\documentclass[useAMS,usenatbib]{mn2e}
\usepackage{graphicx,epsfig}

\newcommand {\hi} {H\,{\small I}}
\newcommand {\fhi}{$F_{\tiny\mbox{H{\sc i}}}$}
\newcommand {\vhi}{$V_{\tiny\mbox{H{\sc i}}}$}
\newcommand {\vopt}{$V_{\tiny\mbox{\it opt}}$}
\newcommand {\nhi}{$N_{\tiny\mbox{H{\sc i}}}$}
\newcommand {\mhi}{$M_{\tiny\mbox{H{\sc i}}}$}
\newcommand {\md}{$M_{\tiny\mbox{dyn}}$}

\newcommand {\lb}{$L_{\tiny\mbox{\it B}}$}

\title[Origin of the Counter-Rotating Gas in NGC 1596]{Origin of the Counter-Rotating Gas in NGC 1596}
\author[Chung et al.]{Aeree Chung$^{1}$\thanks{E-mail:archung@astro.columbia.edu}, 
        B{\"{a}}rbel Koribalski$^{2}$\thanks{E-mail:Baerbel.Koribalski@csiro.au},
        Martin Bureau$^{3}$\thanks{E-mail:bureau@astro.ox.ac.uk} and 
        J. H. van Gorkom$^{1}$\thanks{E-mail:jvangork@astro.columbia.edu}\\
$^{1}$Department of Astronomy, Columbia University, 550 West 120th Street, New York, NY 10027, U.S.A.\\
$^{2}$Australia Telescope National Facility, CSIRO, PO Box 76, Epping, NSW 1710, Australia \\
$^{3}$Sub-Department of Astrophysics, University of Oxford,
Denys Wilkinson Building, Keble Road, Oxford OX1 3RH, United Kingdom}

\begin{document}

\date{Draft Version}

\pagerange{\pageref{firstpage}--\pageref{lastpage}} \pubyear{2002}

\maketitle

\label{firstpage}

\begin{abstract}
We present Australia Telescope Compact Array (ATCA) \hi\ imaging of the edge-on galaxy
NGC~1596, which was recently found to have counter-rotating ionized gas in its center 
($<15''$). We find a large \hi\ envelope associated with a nearby companion, the dwarf
irregular galaxy NGC~1602. The \hi\ covers a region $\approx11\farcm9\times13\farcm4$ 
(62$\times70$ kpc$^2$) and the total \hi\ mass detected is 2.5$\pm0.1\times10^9~M_\odot$
(assuming an 18 Mpc distance). The \hi\ is centered on NGC~1602 but appears to have two 
tidal tails, one of which crosses over NGC~1596. The \hi\ located at the position of 
NGC~1596 has a velocity gradient in the same sense as the ionized gas, i.e. opposite to 
the stellar rotation. Both the existence of a large gas reservoir and the velocity 
gradient of the \hi\ and the ionized gas strongly suggest that the ionized gas in NGC~1596 
originated from NGC~1602. From the length of the \hi\ tails we conclude that the 
interaction started at least 1 Gyr ago, but the unsettled, asymmetric distribution of 
the ionized gas suggests that the accretion occured more recently. NGC~1596 thus provides
a good example where the presence of counter-rotating gas can be directly linked to an 
accretion event. After the accretion has stopped or the merging is complete, NGC~1596 
may evolve to a system with more extended counter-rotating gas but no obvious signature 
of interaction. There is a substantial local \hi\ peak in one of the two tails, 
where we also find a faint stellar counterpart. The \mhi/\lb\ ratio in this region 
is too high for a normal dwarf elliptical or a low surface brightness galaxy, so we
conclude that a tidal dwarf is currently forming there.
\end{abstract}


\section{Introduction}
During the past 30 years evidence has accumulated for the kinematic complexity of a 
significant fraction of ellipticals, lenticulars and some spiral galaxies 
\citep{rubin94}. Oddly rotating stellar and gaseous cores in ellipticals can be 
taken as support that these galaxies are formed in major mergers. A good example is 
the prototypical merger remnant NGC 7252, which has  star-gas counter-rotation in its 
core \citep[e.g.][]{ss98}. More recently, Bertola, Buson and Zeilinger (1992) found 
that, even in a significant fraction of S0 galaxies, the warm and cold gas may be of 
external origin. The first broad search for counter-rotating gas and stars in a sample
of galaxies representative of all morphological types and of the local galaxy luminosity
function \citep{kf01} showed that counter-rotators can be found in S0's over a large
range of luminosity, but that they occur much less frequently among Sa's and are 
absent in later type galaxies. Recently, \citet{drdzh04} found some dwarf
ellipticals with kinematically decoupled cores and concluded that they probably 
originate from an interaction with a giant elliptical or a flyby with a massive galaxy,
while a merger origin is ruled out. Geha, Guhathakurta \& van der Marel (2005)
discovered a counter-rotating core in a low luminosity elliptical and argued for a 
minor merger between two dwarf size galaxies as the most likely origin. 

\begin{figure*}
\includegraphics[width=15cm]{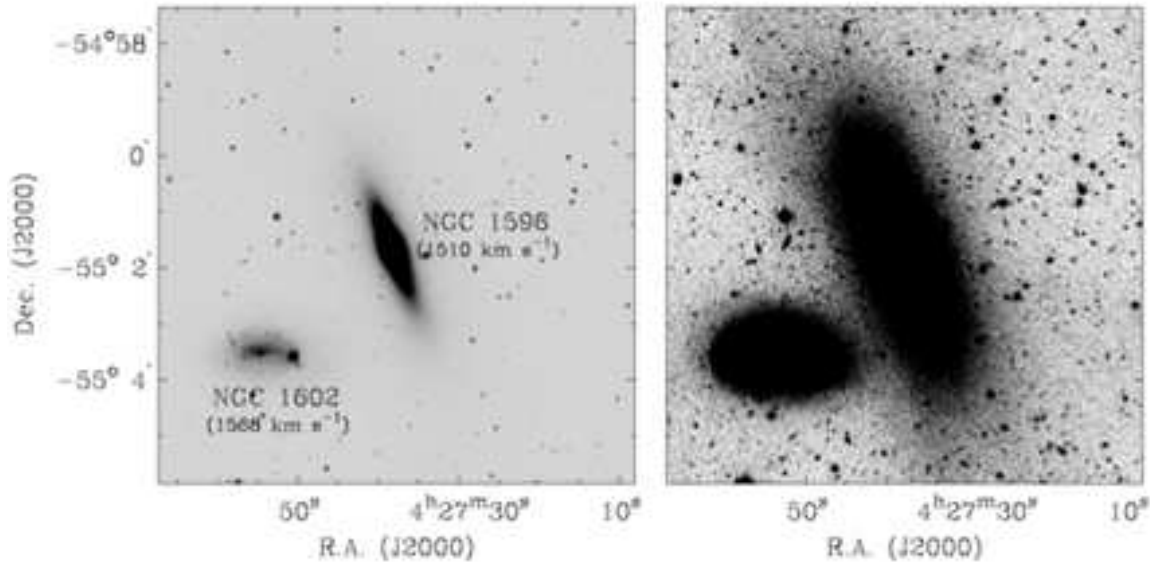}
\caption[]{Optical $R-$band image of NGC~1596 and NGC~1602 \citep{pbld04} in lower 
(left) and higher (right) contrast. The size of the image is $8\farcm5\times8\farcm5$.
NGC~1602 (IB(s)m pec) is the only catalogued galaxy within a 10$'$ radius around 
NGC~1596 and in a similar redshift range ($\Delta V\approx60$ km~s$^{-1}$; see Table 
\ref{tbl-sample}). The projected distance between the two galaxies is $\approx3.0'$ or
15.7 kpc for a distance of 18 Mpc. 
\label{fig-dss}} 
\end{figure*}

\begin{table}
\label{tbl-sample}
\caption{General Properties of NGC~1596 and NGC~1602}
\begin{center}
\begin{tabular}{lcc}
\hline\hline
& NGC 1596 & NGC 1602\\
\hline
\noalign{\vspace{0.1cm}}
$\alpha$ (J2000) & $04^{\rm{h}}27^{\rm{m}}37.8^{\rm{s}}$ 
                 & $04^{\rm{h}}27^{\rm{m}}53.7^{\rm{s}}$ \\
$\delta$ (J2000) & $-55^\circ01'37''$& $-55^\circ03'22''$\\
Morphological type         & SA0: sp    & IB(s)m pec:\\
$D_{25}$ (arcmin)          & 3.71       & 1.94       \\
$m_{_{T}}(B)$ (mag)           & 12.10      & 13.33      \\
\vopt~ (km~s$^{-1}$)$^{a}$  & 1510       & 1568       \\
\fhi~(Jy km~s$^{-1}$)$^{a}$& 15.7       & 11.2       \\
Inclination (deg)$^{b}$    & 90         & 74         \\
Position angle (deg)       & 20         & 83         \\
\hline
\end{tabular}\\
\end{center}

{Refer to \citet{rc3}, unless noted.}\\
$^{a}${\citet{rmgvws82}}\\
$^{b}${Hyperleda ({\it http://leda.univ-lyon1.fr/}). }
\end{table}

A major merger origin for S0 galaxies with counter-rotating gas is thus a possibility,
but other plausible scenarios include minor mergers, accretion of gas from a flyby or 
secondary gas infall (Kannappan \& Fabricant 2001 and references therein). In a recent
spectroscopic study of 30 edge-on disc galaxies, \citet{bc06} reported the discovery 
of counter-rotators in 10\% of their sample or $\approx21\%$ of the S0's only 
\citep[see also][]{cb04, bf99}, consistent with the findings of \citet{bbz92} and 
\citet{kfm96}. NGC~128, NGC~1596 and NGC~3203 revealed ionized gas (mainly [O{\sc iii}])
rotating in the opposite direction to the bulk of the stars. These 3 galaxies have many
properties in common: 1) they are all morphologically S0's, 2) as edge-on systems, they
all show a boxy or peanut-shaped (B/PS) bulge, 3) the ionized gas is only seen within a
$\la$15 arcsec (typically $1.3-4.4$ kpc) region around the center and is highly asymmetric,
and 4) all of them have galaxies nearby in projection on the sky and in a similar redshift
range, except for NGC~3203 whose companions have unknown redshifts. 

The existence of nearby galaxies around these counter-rotators suggests that the origin 
of the counter-rotating gas is external and related to tidal interactions or merger 
events. To find direct evidence of this, we obtained \hi\ follow-up observations of 
these three galaxies. We additionally observed NGC~7332, a similar counter-rotator 
(Fisher, Illingworth \& Franx 1994) recently studied in detail by \citet{fetal04}.

In this paper, we present \hi\ observations of NGC~1596 using the Australia Telescope
Compact Array (ATCA)\footnotemark. \footnotetext{The ATCA is part of the Australia
Telescope which is funded by the Commonwealth of Australia for operation as a National 
Facility managed by CSIRO.} \hi\ observations of the other three systems using the Very
Large Array (VLA) will be presented elsewhere. The redshifts measured in \hi\ are 1510 
km~s$^{-1}$ for NGC~1596 and 1568 km~s$^{-1}$ for NGC~1602, its dwarf companion 
\citep{rmgvws82}. Assuming a distance of 18 Mpc (corrected for Virgocentric infall), 
the projected distance between NGC~1596 and NGC~1602 is 15.7 kpc, less than the distance 
between the Milky Way and the Large Magellanic Cloud. NGC~1596 is about four times brighter 
than NGC~1602 in the $B-$band. It is an S0 galaxy with a B/PS bulge, probably the
edge-on view of a thickened bar \citep{cb04}. NGC~1602 is a dwarf irregular galaxy 
(IB(s)m pec). A deep optical image \citep[Fig. \ref{fig-dss}; ][]{pbld04} reveals 
extended stellar envelopes around both galaxies. \citet{pbld04} argue that the outer 
envelope of NGC~1596 has been disturbed by the interaction with NGC~1602. The general 
properties of the two galaxies are summarised in Table \ref{tbl-sample}. 

In this paper, we will address the following questions:

\begin{enumerate}
\item Are these two galaxies, NGC~1596 and NGC~1602, actually interacting with each 
      other as \citet{pbld04} suggested? What is the evidence for the interaction 
      found in the \hi\ observations?
\item If so, is the interaction responsible for the counter-rotating gas in NGC~1596,
      i.e. does the gas originate from the accreted material?
\item If the origin is accretion, then what is the timescale of the interaction, to go 
      from tidal features to counter-rotating gas?
\item What are the other consequences of the interaction? 
      How does the tidal interaction between NGC~1596 and NGC~1602 influence their
      morphological types (e.g. is it responsible for the bar of NGC~1596 or possibly 
      of NGC~1602)?
\end{enumerate}

Since this study focuses on only one galaxy, we do not present statistics on 
counter-rotators, but will focus instead on providing ideas on how counter-rotation 
between gas and stars can arise. For a discussion of the frequency of counter-rotators,
see the recent paper by \citet{bc06}.

The paper is organised as follows. In section \ref{obsred}, we summarise the observations
and data reduction. The results are then presented and quantified in section \ref{result}. 
We discuss plausible accretion scenarios in section \ref{discuss} and conclude briefly 
in section \ref{conclude}.

\section{Observations and Data Reduction}
\label{obsred}

\begin{table}
\label{tbl-obs}
\begin{center}
\caption{ATCA Observing parameters}
\begin{tabular}{lcc}
\hline\hline
Configurations  & \multicolumn{2}{c}{EW352 (2003 Oct 6$-$7)}  \\
                & \multicolumn{2}{c}{1.5D (2003 Nov 10$-$11)} \\
                & \multicolumn{2}{c}{750A (2004 Feb 20$-$21)} \\
Pointing center (J2000)
& \multicolumn{2}{c}{RA: 04$^{\rm h}$27$^{\rm m}$46.3$^{\rm s}$}\\
& \multicolumn{2}{c}{Dec: $-55\degr$02$\arcmin$29.5$\arcsec$}\\
Primary calibrator & \multicolumn{2}{c}{PKS 1934$-$638} \\
Secondary calibrator & \multicolumn{2}{c}{PKS 0407$-$658} \\
\hline
& \hi\ & 20 cm continuum \\
\hline
Central frequency (MHz) & 1413        & 1384       \\
Total bandwidth (MHz)   & 8           & 128        \\
No. of channels per polarization & 512         & 32         \\
Primary beam (FWHM)     & 33$\farcm$7 & 34$\farcm$4\\
\hline
\end{tabular}
\end{center}
\end{table}

The observations were made using the ATCA on 2003 October 6-7, November 10-11 and 2004 
February 20-21. In order to get good $uv$ coverage the data were obtained using three 
arrays (EW352, 750A and 1.5D), for 12 hr each. \hi\ line and continuum data were obtained 
simultaneously in two (XX,YY) and four (XX,YY,XY,YX) polarizations, respectively. The 
\hi\ line observations were centered at 1413 MHz with an 8 MHz bandwidth and 512 
channels per polarization. This set-up yields $\approx1700$ km~s$^{-1}$ velocity 
coverage and $3.3$ km~s$^{-1}$ per channel. The continuum was centered at 1384 MHz with
a 128 MHz bandwidth and 32 channels per polarization. The pointing was centered between
NGC~1596 and NGC~1602. For further details of the observations, see Table \ref{tbl-obs}.

\begin{figure}
\includegraphics[width=8.5cm]{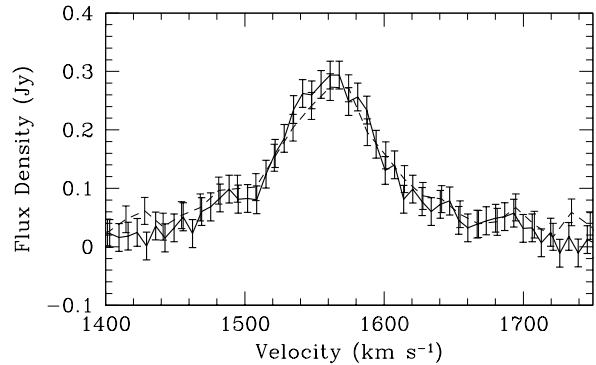}
\caption[]{Global \hi\ spectrum of the galaxy pair NGC1596/1602 as measured with the 
ATCA (solid line; this paper) and the 64-m Parkes telescope \citep[dotted line; see][]
{koribalski04}. Errorbars represent 1$\sigma$.\label{fig-prof}}
\end{figure}

The data were calibrated with the {\tt MIRIAD} (Multichannel Image Reconstruction, Image 
Analysis and Display) package. Antenna gains were derived from the flux of PKS~1934$-$638, 
which was observed at the beginning of each run. Phase variations during the observations 
were corrected using PKS~0407$-$658, which was monitored once every 50 minutes. Bad data 
were manually flagged and the continuum was subtracted using a linear fit through the 
line free channels, as determined from the single-dish \hi\ global profile (HIPASS; 
Koribalski et al. 2004; see Fig. \ref{fig-prof}).

An image cube was generated by combining the data from the three configurations. In order
to maximise sensitivity without unnecessarily degrading the spatial resolution, we applied 
a $uv-$weighting scheme intermediate between uniform and natural (but closer to natural) 
by setting {\tt robust=1} \citep{briggs95}. The \hi\ cube was Hanning-smoothed in velocity 
to a resolution of 6.6 km~s$^{-1}$, cleaned and corrected for primary-beam attenuation.

The beam size of the \hi\ data cube made from the combined data sets is 
$81\farcs9\times65\farcs5$ and the rms noise is 1.9 mJy per beam per 6.6 km~s$^{-1}$ channel. 
This corresponds to a sensitivity limit of 3$\sigma\approx7.7\times 10^{18}$ atoms cm$^{-2}$ 
per channel, which should be sufficient to detect tidal features \citep{hvg96}. A 20 cm 
continuum map was obtained by averaging the continuum channels. The continuum synthesized 
beam is $81\farcs0\times64\farcs6$ and the rms noise is 0.33 mJy beam$^{-1}$. Unless it 
is noted, the \hi\ data presented in this paper are the combined results from the three 
array configurations. \hi\ cubes of each array separately were however also generated. 
The resultant beam sizes are $\approx151''\times126''$, $69''\times49''$ and 
$30''\times24''$ for the individual \hi\ cubes from the EW352, 750A and 1.5D 
configurations, respectively. 

Total \hi\ maps were obtained by summing along the velocity axis using the task {\tt momnt}
in NRAO's Astronomical Image Processing System (AIPS). In order to preserve diffuse and 
extended features, a relatively low cutoff was used and pixels above $\approx1\sigma$ in
a cube smoothed by a factor of 2 spatially and spectrally were used as a mask for the 
full resolution cube. To illustrate the intricate kinematic structure of this system, 
we use channel maps, 3-dimensional rendering of the cube and position-velocity diagrams.

\begin{figure*}
\hfil{\includegraphics[width=0.7\textwidth]{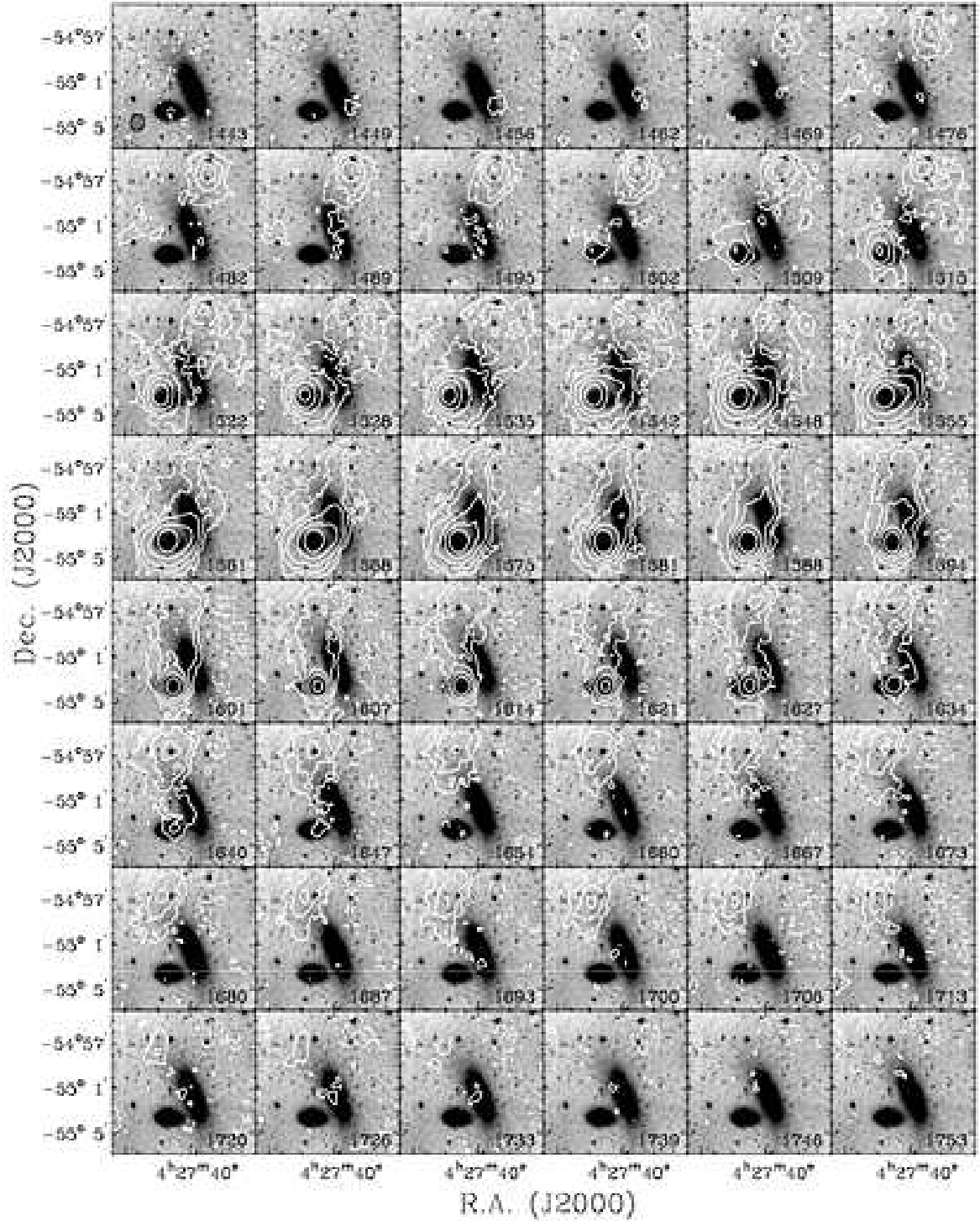}}
\caption[]{\hi\ channel maps of NGC~1596 and NGC~1602 overlaid on a deep optical image 
\citep{pbld04}. The channel maps with the synthesized beam of 89$''\times73''$ are 
presented in order to show the continuity of the entire \hi\ structure more clearly. Note
that the NW tail is connected with NGC~1602 while the NE tail starts from NGC~1602, 
crosses over NGC~1596 and bends to the NE. The heliocentric velocity of each channel is 
indicated at the bottom-right corner of each panel (km~s$^{-1}$). Contour levels are -4, 
-2 (dashed), 2, 4, 8, 16, 32 and 64$\sigma$ (solid) where $\sigma=1$ mJy beam$^{-1}$. 
The synthesized beam is shown at the bottom left of the first panel.
{\bf From top-left to right, $\bf{v=1443}$ km~s${\bf^{-1}}$ with 
${\bf \Delta v\approx6}$ km~s${\bf^{-1}}$.
Channel maps in higher resolution are available through {$\bf http://xxx.lanl.gov/archive/astro-ph$}.}
\label{fig-chdss}}
\end{figure*}
 
\begin{table*}
\begin{minipage}{100mm}
\begin{center}
\label{tbl-region}
\caption{\hi\ Flux densities and masses}
\begin{tabular}{lrrrc}
\hline\hline
\noalign{\vspace{0.1cm}}
                & \multicolumn{1}{c}{\nhi$^{\rm peak}$} 
                & \multicolumn{1}{c}{\fhi} 
                & \multicolumn{1}{c}{\mhi} 
                & \vhi \\
                & \multicolumn{1}{c}{($10^{20}$ cm$^{-2}$)} 
                & \multicolumn{1}{c}{(Jy km~s$^{-1}$)} 
                & \multicolumn{1}{c}{($10^9~M_{\odot}$)}
                & km $s^{-1}$\\
\noalign{\vspace{0.1cm}}
\hline
\noalign{\vspace{0.1cm}}
NW tail  &  3.1~~~~~~~&  6.3~~~~~~ & 0.48~~~ & $1462-1528$\\
NGC~1602 & 10.0~~~~~~~& 19.1~~~~~~ & 1.46~~~ & $1495-1647$\\
NGC~1596 &  $-$~~~~~~~&  3.3~~~~~~ & 0.25~~~ & $1522~?-1647~?$\\
NE tail  &  1.8~~~~~~~&  4.6~~~~~~ & 0.35~~~ & $1548-1733$\\
Total    & 10.0~~~~~~~& 33.0~~~~~~ & 2.52~~~ & $1462-1733$\\
\noalign{\vspace{0.1cm}}
\hline
\end{tabular}
\end{center}
\end{minipage}
\end{table*}

\begin{figure*}
\includegraphics[width=8cm]{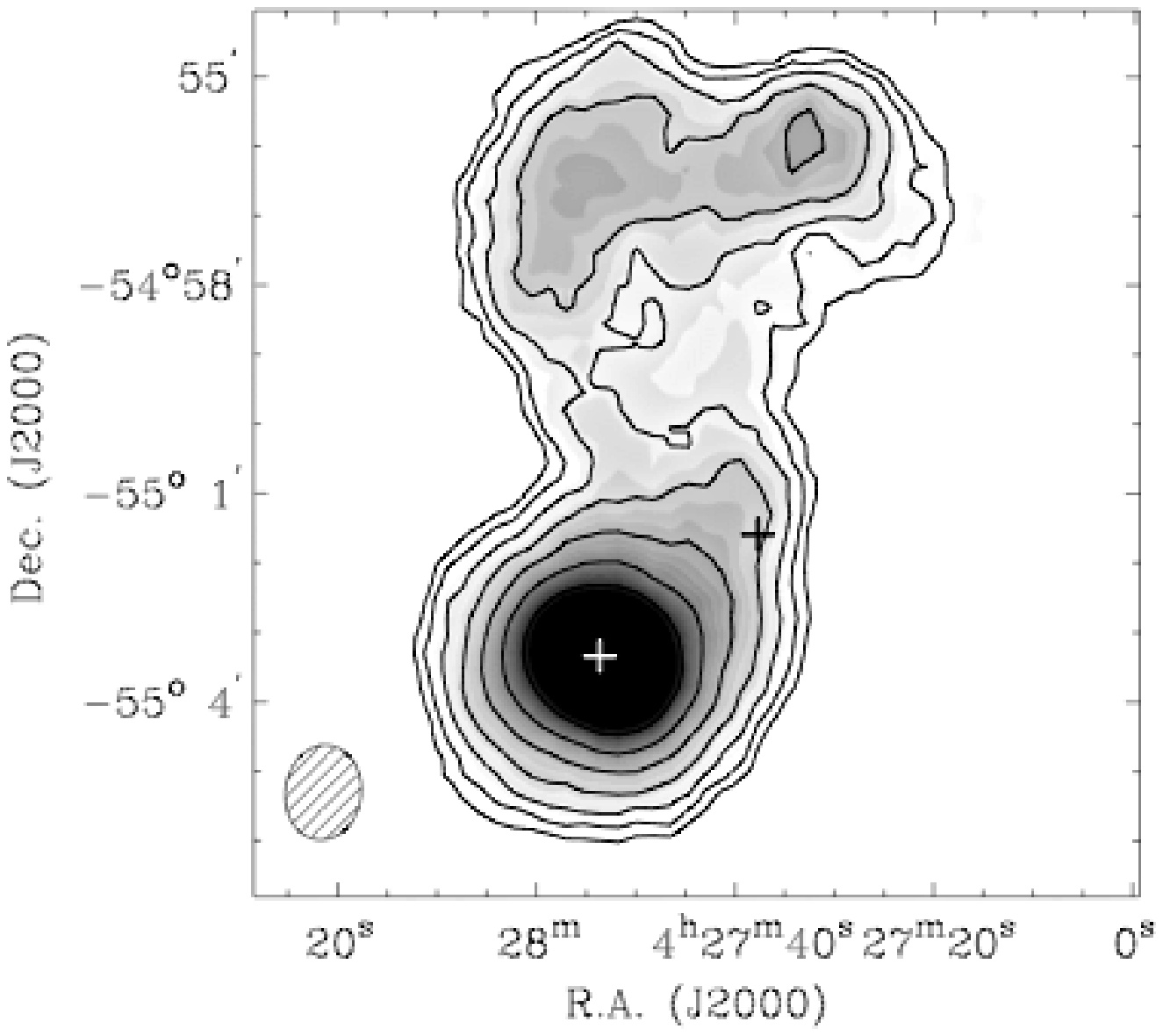}
\includegraphics[width=8cm]{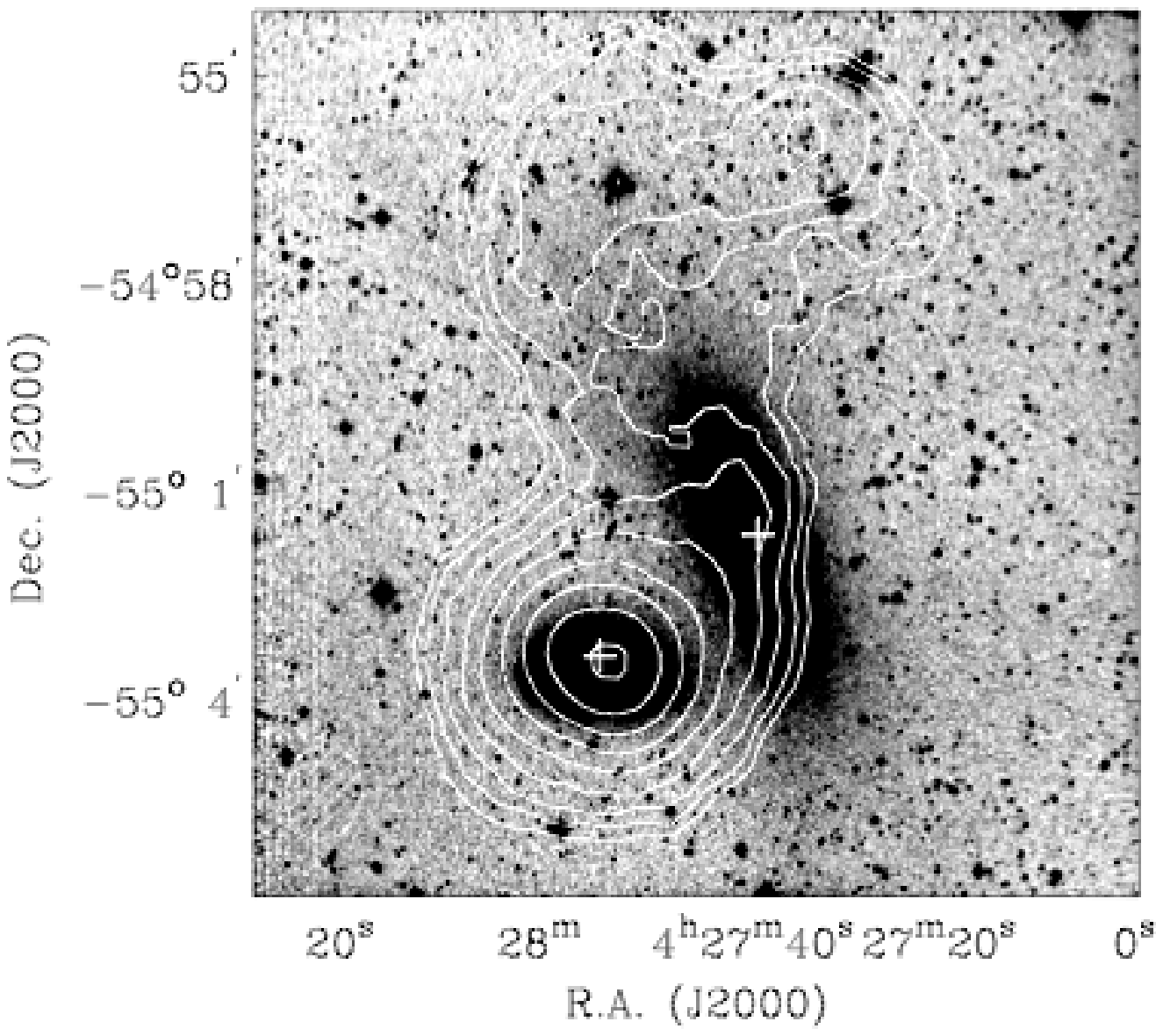}
					 
\includegraphics[width=8cm]{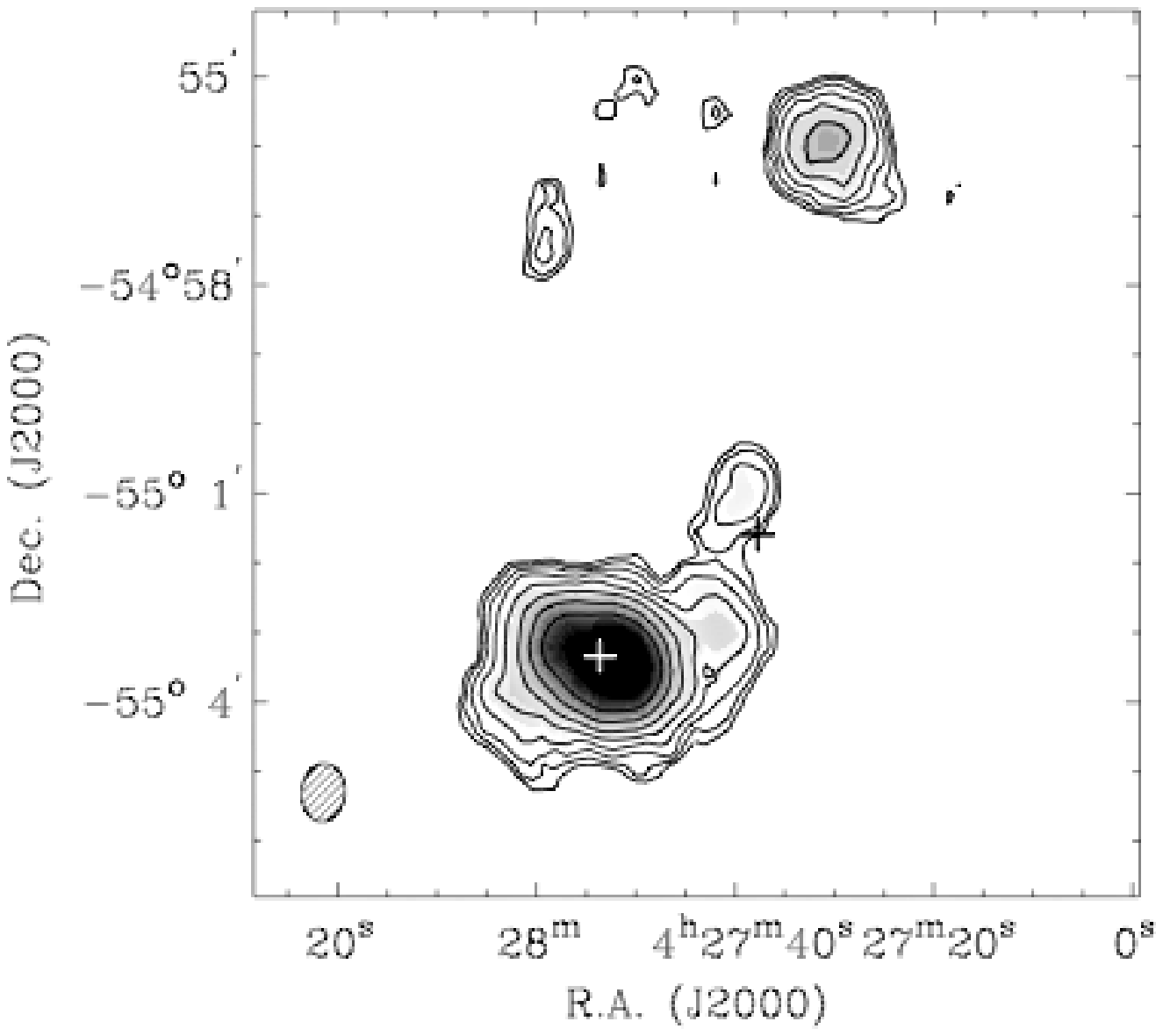}
\includegraphics[width=8cm]{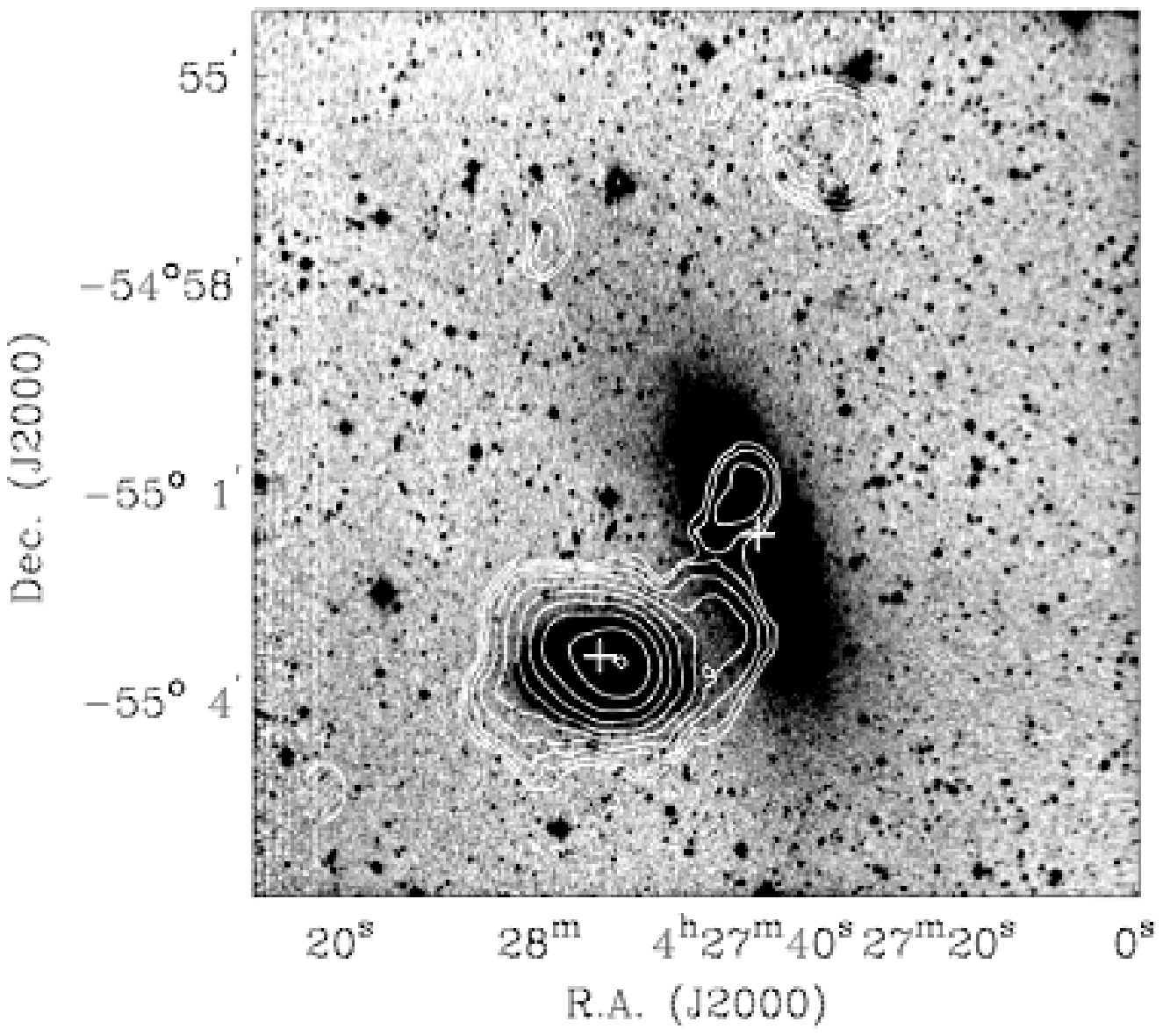}
\caption[]{Total \hi\ maps. Top) Total \hi\ distribution overlaid on a grayscale 
representation (left) and a deep optical image \citep[right;][]{pbld04}. Contour levels 
are 4, 6, 9, 13.5, 20.3, 30.5, 45.7, 68.6 and 102.8$\times10^{19}$ cm$^{-2}$ (0.2, 0.3, 
... Jy km~s$^{-1}$ beam$^{-1}$). Bottom) Higher resolution \hi\ map. Contours are 4, 6, 
9, 13.5, 20.3, 30.5, 45.7, 68.6, 102.8 and 154.2$\times10^{19}$ cm$^{-2}$ (0.07, 0.10, 
... Jy km~s$^{-1}$ beam$^{-1}$), with a synthesized beam of $49\farcs9\times37\farcs6$.
Crosses indicate the optical centers of the two galaxies and the synthesized beams are 
shown at the bottom left of each panel.\label{fig-mom0}}
\end{figure*}

\section{Results}
\label{result}

\subsection{\hi\ distribution and kinematics}

\hi\ is found in a broad region around NGC~1596 and NGC~1602 with an extension of 
$11\farcm9$ N-S,  $13\farcm4$ E-W and a velocity range of 250 km~s$^{-1}$ (see Figs. 
\ref{fig-prof}-\ref{fig-mom0}). Assuming an 18 Mpc distance, this corresponds to 
$\approx62\times70$ kpc$^2$. Most of the gas is concentrated around NGC~1602, which is
clearly the main and presumably the original reservoir of \hi. The extended \hi\ appears
to show two tidal tails, one to the NNW and the other one to the NNE from NGC~1602, at 
lower and higher velocities respectively (hereafter NW tail and NE tail). The \hi\ total
flux of 33.0$\pm$1.3 Jy~km~s$^{-1}$ seen by ATCA is in agreement with the flux of 
36.5$\pm$4.4 Jy~km~s$^{-1}$ found by HIPASS \citep{koribalski04}, so the interferometer
appears to have detected most of the extended \hi\ emission (Fig. \ref{fig-prof}).

Total \hi\ maps are presented in Figure \ref{fig-mom0}. The region around NGC~1602 contains
$\approx$58\% of the total flux (19.1 Jy km~s$^{-1}$ or \mhi$\approx1.5\times10^9~M_\odot$).
The structure is highly asymmetric, the \hi\ stretching in the direction of NGC~1596. Based
on the highest resolution map (beam $\approx30''\times24''$), the peak in \hi\ emission is
offset to the west by $\approx29''$ compared to NGC~1602's optical center. In Figure 
\ref{fig-mom0}, we additionally present the total \hi\ map generated from the two high 
resolution data sets only (1.5D and 750A). The offset of the \hi\ distribution from the
optical center and the extended outer envelope toward NGC~1596 appear more clearly in 
total \hi\ maps of higher resolutions.

The \hi\ surrounding NGC~1596 is far from that of normal spirals. About 10\% of the total
flux (3.3 Jy km~s$^{-1}$ or \mhi$\approx2.5\times10^8~M_\odot$) is found within an ellipse
of $2D_{25}$ major-axis centered on NGC~1596, without any distinctive concentration. The 
peak column density within the optical radius of NGC~1596 is about $1.4\times10^{20}$ 
cm$^{-2}$. 

The properties of these distinct regions are summarised in Table 3.

The NW tail is quite distinct from the rest of the \hi\ structure both kinematically and 
morphologically. It extends from $\approx$1462 to 1528 km~s$^{-1}$ (Fig. \ref{fig-chdss}), 
containing $\approx$19\% of the total flux (6.3 Jy km~s$^{-1}$ or 
\mhi$\approx4.8\times10^8~M_\odot$). The \hi\ emission peaks at 
$\alpha\approx4^{\rm h}27^{\rm m}30^{\rm s}$, $\delta\approx-54^{\circ}56'00''$ with a column 
density of $3.1\times10^{20}$ cm$^{-2}$. Note that the deep optical image shows very faint 
emission at the tip of the NW tail (near $\alpha\approx4^{\rm h}27^{\rm m}31.5^{\rm s}$,
$\delta\approx-54^{\circ}55'54''$), coincident with the \hi\ peak. This may in fact be a 
tidal dwarf as discussed in section 4. The tail can be most clearly seen extending from 
the dwarf to NGC~1602 in channels, 1469-1509 km~s$^{-1}$.

The rest of the structure, i.e. the NE tail, extends from 1548 to 1733 km s$^{-1}$ (Fig. 
\ref{fig-chdss}) and contains about 14\% of the total flux (4.6 Jy km~s$^{-1}$ or 
\mhi$\approx3.5\times10^8~M_\odot$). The peak emission is not so different, with a highest 
column density of 1.8$\times10^{20}$ cm$^{-2}$ ($\alpha\approx4^{\rm h}28^{\rm m}00^{\rm s}$, 
$\delta\approx-54^\circ55'00''$). 

\begin{figure*}
\begin{center}
\hfil{\includegraphics[width=1.0\textwidth]{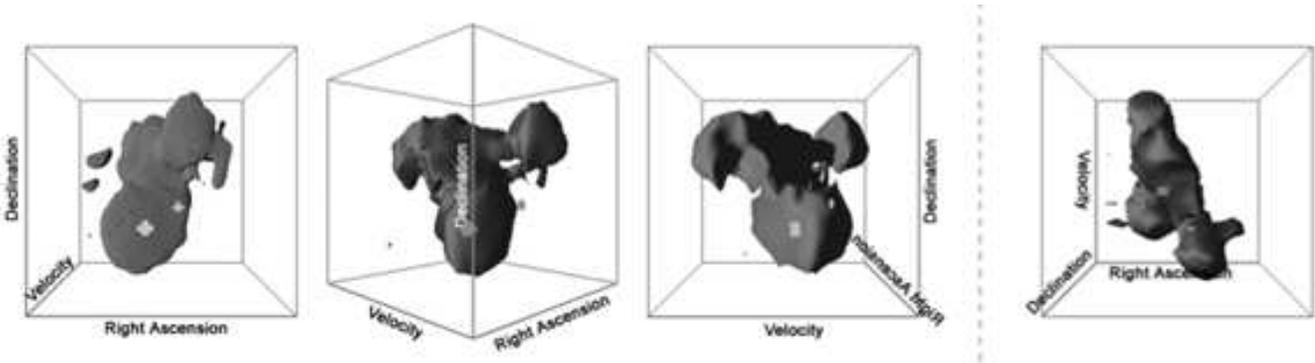}}
\caption[]{Three-dimensional view of the \hi\ emission in NGC~1596 and NGC~1602. 
The first three from left to right are a view along our line of sight, with heliocentric 
velocity on the third axis, and similar views rotated by 45$^\circ$ and 90$^\circ$.
The figure on the rightmost, a view from the top is shown. NGC~1596 and NGC~1602 are 
indicated with crosses.
\label{fig-3dgas}}
\end{center}
\end{figure*}

\subsection{20 cm radio continuum}

We do not detect NGC~1596, NGC~1602 or the tidal tails in the continuum down to a 
3$\sigma$ limit of 1 mJy beam$^{-1}$ (a star formation rate of 0.09 $M_\odot$~yr$^{-1}$;
Hopkins et al. 2003). A prominent source (PMN~J0427-5506) is however located just south 
of NGC~1596 and NGC~1602, with a peak of 0.24 Jy per beam 
($\alpha\approx4^{\rm h}27^{\rm m}40^{\rm s}$, $\delta\approx-55^\circ07'00''$). Our 
measurement at 1.4 GHz fits nicely in the middle of previous measurements at 4.85 GHz 
\citep[107$\pm9$ mJy;][]{wgbe94} and 0.843 GHz \citep[466$\pm$14 mJy;][]{mauch03}. 
No \hi\ absorption was detected against PMN~J0427-5506.

\section{Discussion}
\label{discuss}

\subsection{Interaction Scenario}

The \hi\ morphology and kinematics suggest that some \hi\ gas has been tidally stripped 
from NGC~1602 by NGC~1596. 

First, two arms (the NW and NE tails) are located at velocities lower and higher than NGC~1602,
suggesting that they point in opposite directions with respect to each other along the 
line-of-sight, similarly to the features formed in tidal interactions \citep[e.g.][]{tt72}. 

\begin{figure}
\begin{center}
\includegraphics[width=7cm]{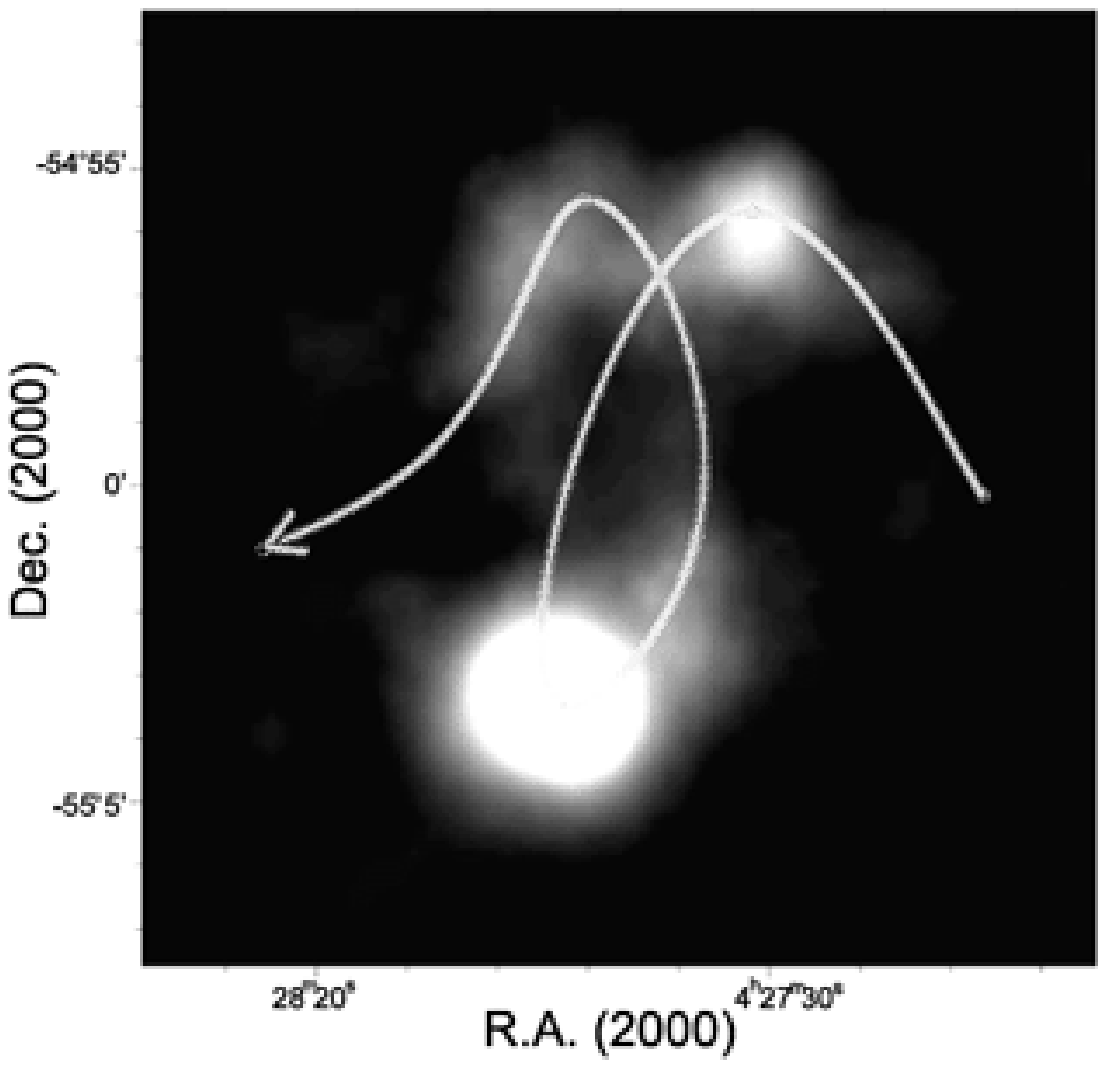}

\includegraphics[width=7cm]{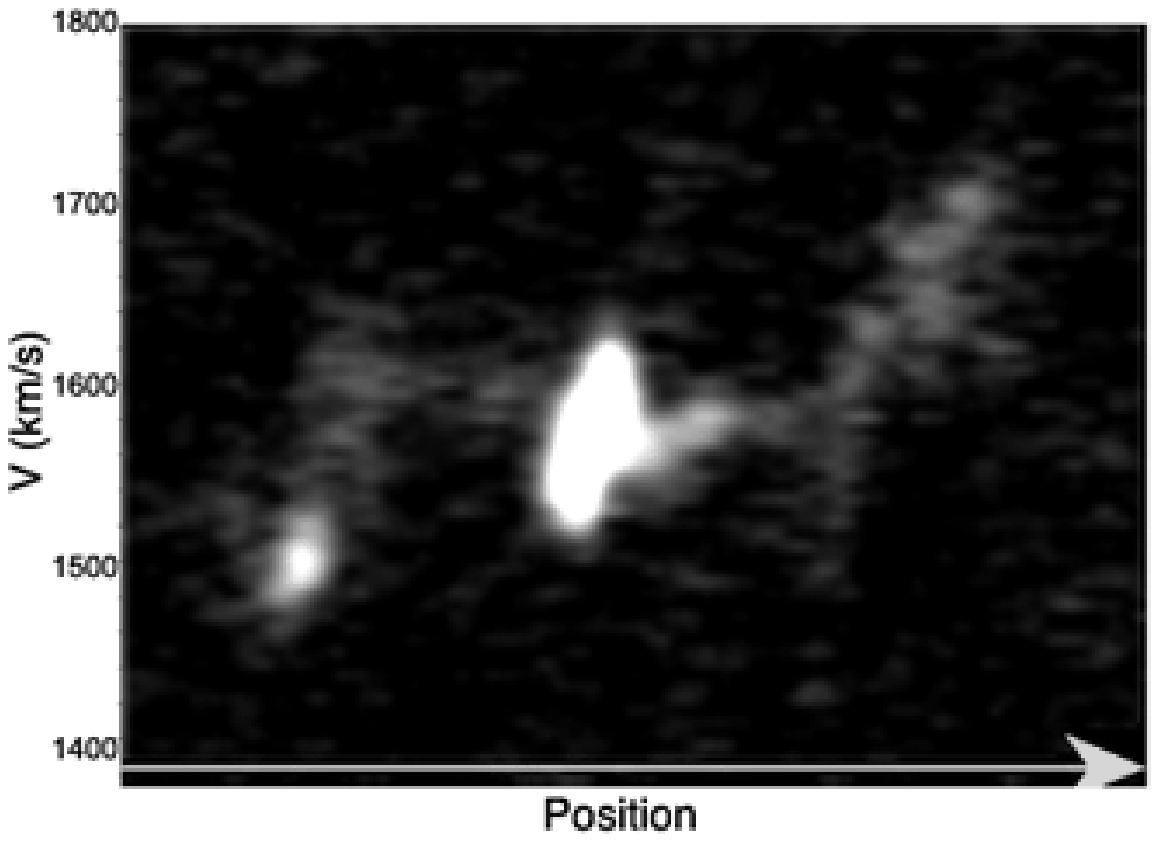}
\caption[]{A position-velocity cut along the \hi\ structure, from the NW tail through 
NGC~1602, NGC~1596 and the NE tail, as indicated by the line with an arrow. The width 
of the 
cut is 15$''$ and the total length of the cut $\approx32.5'$ (upper panel). The NW 
tail (lower velocities) is not only spatially but also kinematically more distinct 
than the NE tail. The cut clearly shows that the \hi\ around NGC~1602, the major 
\hi\ reservoir, is highly asymmetric and more extended toward NGC~1596.
\label{fig-spcut}}
\end{center}
\end{figure}

Second, the complex \hi\ morphology strengthens the suggestion made by \citet{pbld04} 
that NGC~1596 and NGC~1602 have interacted in the past. NGC~1602 most likely had a very 
extended \hi\ disc, as is common for dwarf irregular galaxies \citep[e.g.][]{hetal98}. One or 
several close passages of NGC~1602 near the \hi\ poor S0 galaxy NGC~1596 pulled out what
appears to be two \hi\ tidal tails. The two tails can best be seen in Figures 
\ref{fig-3dgas} and \ref{fig-spcut}. In Figure \ref{fig-3dgas}, we present 3D renderings 
of the \hi\ cube, indicating where NGC~1596 and NGC~1602 are in the plane of the sky and 
along our line-of-sight. Note that the tails are very distinct in velocity. Figure 
\ref{fig-spcut} shows a position-velocity cut along the two tails. The continuous velocity 
gradient along both tails is remarkable, though there is a kink in velocity between the
NW tail and NGC~1602. This kind of jump has been seen before in the tidal tail of a merger
remnant \citep[NGC~7252; ][see their Fig. 1]{hm95}, and  could be the combined effect
of a projection and the tidal tail detaching itself, forming a dwarf galaxy.

Since the tails are in different orbital planes 
(they would otherwise cross), more than one close passage must have taken place. Indeed,
\citet{sl00} showed in their N-body model of M51 that significant out-of-plane velocities
can be produced through multiple encounters. We speculate that the NW tail was pulled out
in a first encounter, while the NE tail was pulled out in a second closer passage. The 
asymmetric \hi\ distribution in NGC~1602 (toward NGC~1596) also suggests that this is the 
most recent event (Figs. \ref{fig-mom0} and \ref{fig-3dgas}). The higher resolution \hi\ 
image of NGC~1602 even shows a hint of a counter tail in the east. 

Third, NGC~1596's stripping of the \hi\ gas from NGC~1602 is plausible in terms of 
timescales. Assuming that the total \hi\ found in this region used to belong to NGC~1602,
roughly 0.9$\times10^9~M_\odot$ of \hi\ has been stripped. The NW and NE tail are about 
49 kpc and 38 kpc long, respectively, projected on the sky (from the center of NGC~1602 
to where those concentrations are present; see section \ref{result}). The \hi\ kinematics tells 
us that it would take roughly $3-4\times10^8$ yr for the gas to travel from the center 
of NGC~1602 to where the tails reach at present. Since there must have been at least two 
passages, this yields a lower limit of about 1 Gyr to form both tails. Some simulations 
\citep[e.g. ][]{hm95} show that it is possible to form tails with those lengths in such 
time scales. Note that the \hi\ mass found in tidal features can be as large as 
$4.5\times10^9~M_\odot$ \citep[e.g. in the Southern tail of NGC~3921; ][]{hvg96}.
These properties suggest that most of the gas originally belonged to NGC~1602 and that 
some of the \hi\ has been stripped by the interaction with NGC~1596, forming the two 
tidal tails.

\subsection{Origin of the Counter-Rotating Ionized Gas}

The question remains whether the stripped \hi\ is responsible for the counter-rotating 
ionized gas found in NGC~1596.

First, the fact that there is a large gas reservoir available around NGC~1596  
makes the connection between the \hi\ and the counter-rotating ionized gas more likely
than an internal origin.

\begin{figure*}
\hfil{\includegraphics[width=0.9\textwidth]{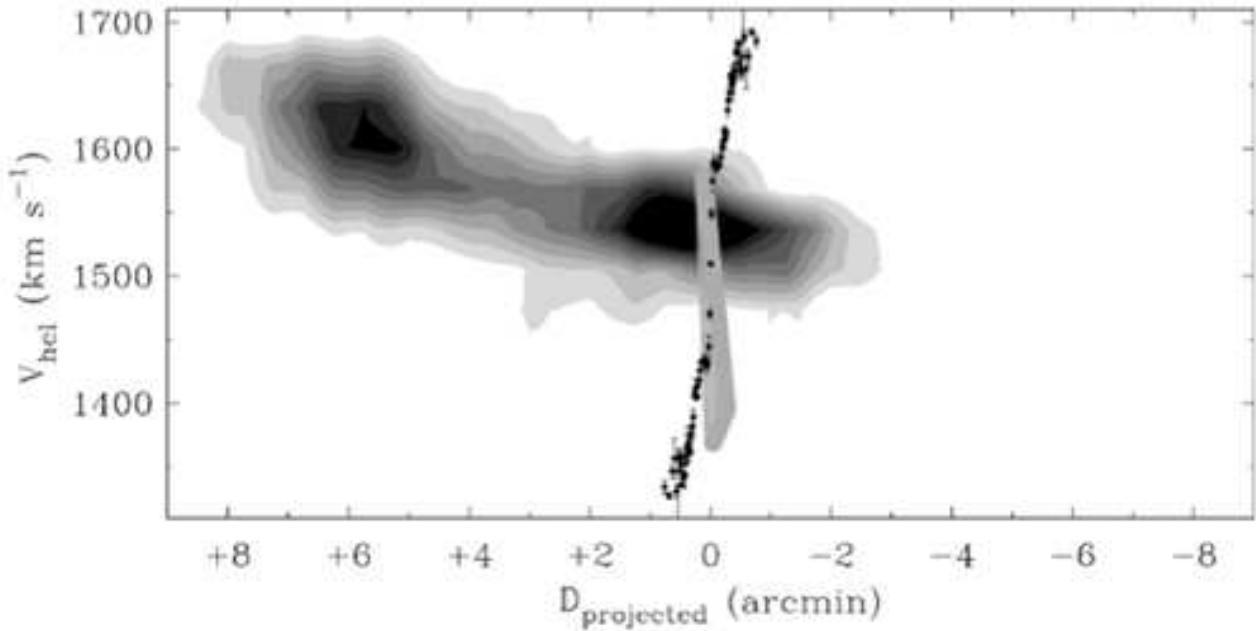}}
\caption[]{The \hi\ position-velocity diagram (PVD) along the major axis (PA=20$^\circ$)
of NGC~1596 (grayscale). The width of the cut is 15$''$. The ionized gas PVD (light gray) and 
stellar rotation curve \citep[black dots;][]{cb04} are overlaid. The leftmost and the rightmost
corners in x-axis correspond to the NE and the SW on the sky, respectively. Note that the ionized 
gas is not the actual data but a schematic representation based on \citet{bc06}. \label{fig-comxv}}
\end{figure*}

In addition, the \hi\ and the ionized gas have the same velocity gradient, opposite to 
that of the stars. In Figure \ref{fig-comxv}, the 
\hi\ position-velocity diagram (PVD) along the major-axis of NGC~1596 is compared with that 
of the ionized gas found in the optical spectra and the stellar rotation curve. The \hi\ 
concentration to the NE and over a large velocity range is the NE tail. It is clear that
the \hi\ and the ionized gas have the same velocity gradient, both rotating in the opposite 
direction to the stars. Whether the accreted gas is co- or counter-rotating with the disc 
clearly depends on how it falls in. However, the large-scale agreement between the velocity 
gradient of the \hi\ and the ionized gas in the disc of NGC~1596 (Fig. \ref{fig-comxv}) makes 
the connection between the \hi\ gas of NGC~1602 and the ionized gas in NGC~1596 more plausible.
The counter-rotating ionized gas in NGC~1596 very likely originated from the \hi\ gas stripped 
from NGC~1602, which must have been accreted in retrograde motion with respect to the stellar 
disc. 

Given the highly asymmetric distribution of the ionized gas in the disc of NGC1596, 
there must not have been enough time for the accreted material to totally settle. 
The accretion onto NGC~1596 must have been ongoing for not much longer than 
$\sim4\times10^8$ yrs, i.e. one rotation period. 

\subsection{Other Consequences of the Interaction}

There is a local \hi\ peak in the NW tail which also shows a faint stellar counterpart
($\alpha_{_{\rm NW}}=4^{\rm h}27^{\rm m}31.5^{\rm s}$, $\delta_{_{\rm NW}}=-54^{\circ}55'54''$).
The peak gas column density of this region in fact exceeds the standard star formation threshold
\citep{kennicutt89} and \citet{fs90} identified a distinct object in this location, 
classifying it as dwarf elliptical ($D=10.9''$ and $B_T=19.8$ mag). This object is however not
likely to be a normal dwarf elliptical, based on its high gas-to-light ratio. The \hi\ mass within 
$R<1.5'$ around ($\alpha,\delta$)$_{_{\rm NW}}$, where the kinematics is fairly well defined (Fig.
\ref{fig-spcut}), is 4.8$\times10^8~M_\odot$. This yields \mhi/\lb$\approx79$ $M_\odot/L_\odot$
(or, assuming a linewidth of 60$-70$ km s$^{-1}$, \md/\lb$\approx$164 $M_\odot/L_\odot$), 
which is much larger than that of typical dE galaxies \citep[$<0.2$, e.g.][]{gj83,bjco05}. 
Also, since gas rich low surface brightness galaxies have \mhi/\lb\ and \md/\lb\ in the range of
$0.3-6$ and $20-75$ $M_\odot/L_\odot$, respectively \citep{freeman97}, it is also unlikely that
the object is an LSB galaxy.

This faint optical counterpart is thus more likely to be a dwarf formed at the end of one 
tidal arm than a normal dE or an LSB that was there before the interaction started. The idea
of creating dwarf galaxies in collisions of galaxies (tidal dwarfs or TDGs) was first proposed
by \citet{zwicky56} and a fair number of candidates have been found (e.g. in NGC~4038/9, 
{\lq\lq}the Antennae{\rq\rq}; Hibbard et al. 2001 and references therein). Deep H$\alpha$/$UV$
imaging or metallicity measurements should enable us to sort out this issue.

As shown in Figure \ref{fig-3dgas}, the tidal tails are not in the same orbital plane, 
implying that there must have been more than one close approach between the two galaxies. 
It is possible that NGC~1596 was approached from the back side of NGC~1602 along the 
line-of-sight, pulling out the NW tail first and the NE tail more recently. This time 
sequence is also consistent with the fact that only the NW tail shows a stellar counterpart. 
The stars could have formed in situ in the NW tail, or alternatively they may have been 
stripped with the gas in the tidal encounter. Further optical observations (e.g. H$\alpha$ 
imaging or spectroscopy) should also help to sort this out. Some of the gas from the NE tail 
must have been accreted onto NGC~1596 (in retrograde motion with respect to the stars) 
after the second approach. 

It is worth recalling that NGC~1596 possesses a boxy bulge, one of the morphological 
signatures of a thickened bar seen edge-on \citep[e.g.][]{cs81,cdfp90}. NGC~1596 also revealed 
kinematic signatures of a bar in the study by \citet{cb04}. The bar may have contributed 
to transport the material to the central part of the disc, although such processes are 
inefficient for counter-rotating material. It is also possible that the interaction with 
NGC~1602 accelerated the formation of the bar in NGC~1596 (or vice-versa) as has been 
shown to occur in simulations \citep[e.g.][]{noguchi87,gca90,mwhmb95,mn98}. 

\section{Conclusion}
\label{conclude}

We presented the \hi\ distribution and kinematics of the galaxy pair NGC~1596/NGC~1602 and
its surrounding. \hi\ is found in a broad region $\approx11\farcm9\times13\farcm4$ 
covering NGC~1596 and its \hi-rich companion NGC~1602. This region corresponds to 
$\approx62\times$70 kpc$^{2}$ assuming an 18 Mpc distance. The total \hi\ mass found is 
2.5$\pm0.1\times10^9~M_\odot$ and about half of this is concentrated in and around NGC~1602. 
Both the \hi\ morphology and kinematics suggest that NGC~1602 is the main gas reservoir.
Two tidal tails are found, respectively at lower and higher velocities with respect to 
NGC~1602, which were most likely caused by an interaction with NGC~1596. The 
counter-rotating ionized gas found in NGC~1596 thus probably originated from  \hi\ gas
transferred from NGC~1602 to NGC~1596. Besides the fact that there is a huge gas reservoir
available in this region, this is also supported by the consistent velocity gradient
of the \hi\ and the counter-rotating ionized gas along the major-axis of NGC~1596.

We find a local \hi\ peak in the NW tail, where we also see a faint stellar counterpart. 
The \mhi/\lb\ (or \md/\lb) ratio of this region is unrealistically high for a normal dE 
or LSB galaxy, and it is more likely to be a dwarf that has formed at the tip of a tidal
 arm (TDG). 

Our discovery of counter-rotating gas in the sample of \citet{cb04} was somewhat fortuitous.
In spite of the difficulty to find such objects, they are known to be common (Bureau \& Chung
2006 and references therein), and although there are only a few known counter-rotators showing
direct evidence for ongoing accretion \citep[e.g. IC~1459,][]{som02}, signatures of past 
mergers or hints of tidal interactions are found in a number of them \citep{schweizer98}. 
Counter-rotators thus support the idea that tidal interactions and accretion play an important
role in galaxy formation and evolution.

\section*{Acknowledgments}
We would like to thank Jim Caswell, Meryl Waugh, Lister Staveley-Smith and Emma Ryan-Weber 
for their technical advice during the observations and data reduction. We thank John Hibbard,
Renzo Sancisi and Chris Mihos for their inspiring comments on this work, and the anonymous
referee for comments that led to an improvement of the paper. 
We also thank Michael Pohlen who kindly provided the deep optical image of NGC~1596/1602. 
This work has been supported in part by NSF grant AST-00-98249 to Columbia University.

\label{lastpage}

\end{document}